\documentstyle[pra,multicol,aps,epsf]{revtex}
\draft
\begin{document}
\title{Gain Components in Autler-Townes Doublet from Quantum Interferences In
Decay Channels}
\author{Sunish Menon$^{1}$  and G. S. Agarwal$^{1,2}$}
\address{$^1$ Physical Research Laboratory, Navrangpura, Ahmedabad 380 009, 
India\\
$^2$ Max-Plank-Institut f\"{u}r Quantenoptik, 85748 Garching, Germany.}
\date{\today}
\maketitle
\begin{abstract}
We consider {\bf non-degenerate} pump-probe spectroscopy of V-systems under conditions such that
interference among decay channels is important.   
We demonstrate how this interference can result in new gain features
instead of the usual absorption features.  We relate this gain to the 
existence of a new vacuum induced quasi-trapped-state.  We further show how this 
also results in large refractive index with low absorption. 
\end{abstract}
\begin{multicols}{2}
\section{introduction}
The properties of a medium are significantly altered when the medium is 
driven by strong, resonant, coherent fields.  Mollow first studied in detail the
physical characteristics of a two-level system driven by a coherent field 
of arbitrary strength.  He discovered new features in the emission spectra
[1a].  These new features are best understood 
in terms of
field dependent eigenstates and eigenvalues or dressed states 
of the system \cite{cohenTbook}.  Mollow in a later paper [1b], further demonstrated
the possibility of amplification of a probe field in a coherently driven
two-level system.  This gain can also be understood in terms of dressed
states \cite{gsa79-cohen93}.
An alternative way of probing a driven two-level model is by coupling a
 probe field to one of the states (e.g. the ground state) of the 
strongly driven transition and 
 a different excited state (V-system).
 The absorption spectra will show new resonances related to the dressed states
of the strongly driven two-level atom.  The strong drive splits the absorption
resonance into two components known as 
the Autler-Townes components.   
Various experiments in gases \cite{gases} and
in solid state systems \cite{solid} have confirmed the
presence of the Autler-Townes splitting of absorption lines.  
 Such a 
three-level model is not known to show gain unless additional fields 
are introduced.  For example, an incoherent pumping along the probing
transition of a V-system will give rise to gain \cite{{xiao94},{zibrov95}}
. Other three-level
models with coherent and incoherent pumping also exhibit gain \cite{olga88}.

In this paper, we demonstrate that quantum interference between different paths
of spontaneous emission  in a V-system can produce {\em gain under 
conditions when one would have otherwise observed absorption peaks}.
Interference due to spontaneous emission can arise when
spontaneous emission from
one level can strongly affect a neighboring transition.  For example, consider
excited levels $|1\rangle$, $|2\rangle$ of same parity, and ground state
$|3\rangle$ of a different parity (see Fig. \ref{one}(a)).  Let the spontaneous
emission rates from levels $|1\rangle$ and $|2\rangle$ to level $|3\rangle$ 
be denoted as $2\gamma_1$ and $2\gamma_2$ respectively.  In interaction picture,
the equations of motion for density
matrix elements    
will be \cite{gsabook}
\begin{eqnarray}
\dot{\rho}_{11} &=& -2\gamma_1 \rho_{11} - \sqrt{\gamma_1\gamma_2}\cos\theta 
(\rho_{12}e^{-iW_{12} t} + \rho_{21}e^{iW_{12} t}),\nonumber \\
\dot{\rho}_{22} &=& -2\gamma_2 \rho_{22} - \sqrt{\gamma_1\gamma_2}\cos\theta 
(\rho_{12}e^{-iW_{12} t} + \rho_{21}e^{iW_{12} t}),\nonumber \\
\dot{\rho}_{12} &=& -(\gamma_1+\gamma_2)\rho_{12} -\sqrt{\gamma_1\gamma_2}
\cos\theta e^{iW_{12}t}(\rho_{11}+\rho_{22}), \nonumber \\
\dot{\rho}_{13} &=& -\gamma_1 \rho_{13} - \sqrt{\gamma_1 \gamma_2}\cos\theta 
e^{iW_{12} t}\rho_{23}, \nonumber \\
\dot{\rho}_{23} &=& -\gamma_2 \rho_{23} - \sqrt{\gamma_1 \gamma_2}\cos\theta 
e^{-iW_{12} t}\rho_{13}.
\label{density0}
\end{eqnarray} 
Here $\hbar W_{12}$ is the energy separation between the excited levels which 
we keep arbitrary.  The
above equations are derived without making any kind of secular approximation, 
and can be solved under the conditions $\rho_{33} + \rho_{22} + \rho_{11} = 1$ 
and $\rho_{ij} = \rho^*_{ji}$.
The off-diagonal radiative coupling terms in the equations for diagonal
elements of $\rho$ are due to interference
among decay channels.  Here the parameter $\theta$ is the angle between the dipole
matrix elements $\vec{d}_{13}$ and $\vec{d}_{23}$, where $\vec{d}_{i3} = 
\langle i|{\bf d}|3\rangle$ ($i = 1,2$) and {\bf d} is the dipole moment 
operator.  Note that this interference exists only when
$\theta \neq 90^o$.   Moreover, for equations (\ref{density0}) when
 $W_{12} \gg \gamma_1, \gamma_2$, the oscillatory terms will average out and
the effects of such off-diagonal terms will vanish.  A result of such an off-diagonal
radiative coupling is that 
the coherence $\rho_{12}(t)$ will
evolve even when $\rho_{12} (0) = 0$.  The coherence arises from the  vacuum
of the electromagnetic field.  We refer to it as vacuum induced coherence (VIC).
This coherence term will change the steady 
state response of the medium and under suitable conditions can
{\em create trapping in a degenerate V-system} \cite{gsabook}.
The coherence can also modify significantly the emission spectrum of a
near-degenerate V-system \cite{agassi84}.  Recent works \cite{russian99} generalize equations
(\ref{density0}) to include thermal photons as well as incoherent pumping.  The presence of both thermal
photons and VIC leads to additional features in the spectrum.

In the present paper, we study how to probe VIC by using a pump-probe 
spectroscopy.  We demonstrate that the VIC can manifest itself via gain 
features instead of the traditional absorption features.
The organization of this paper is as follows: In Sec. II we present basic
equations describing the pump-probe spectroscopy under conditions when 
interference between decay channels is important.  We also point out the crucial 
difference between the present work  and the previous studies.
In Sec. III we discuss our numerical results.
We show the possibility of new gain features  
due to VIC.  In Sec. IV we report the 
possibility of quasi-trapped-states that arise strictly 
from the interference
between decay channels.  In Sec. V  we analyze
the effect of this trapping on the absorption and dispersion properties 
of the pump field. 
In Sec. VI we explain the numerical results of Sec. III in terms of the
trapped states discussed in Sec. IV.  

\section{Basic Equations}
 
Consider the pump-probe set-up shown in Fig. 1(a).  
The transition dipole moments $\vec{d}_{13}$ and $\vec{d}_{23}$ are
non-orthogonal: so in principle we should include the coupling of pump (probe)
to the transition $|1\rangle \leftrightarrow |3\rangle$ ($|2\rangle 
\leftrightarrow |3\rangle$).     
\begin{figure}[h]
\epsfxsize 3.2 in
\epsfysize 1.4 in
\epsfbox{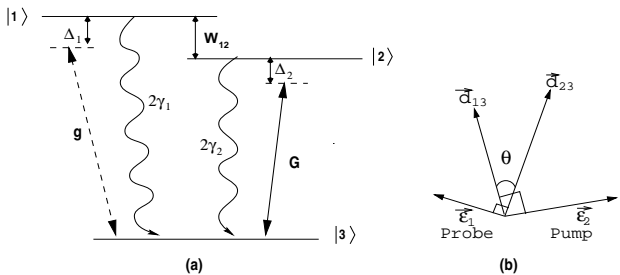}
\vspace*{0.5 cm}
\narrowtext
\caption{(a) Schematic diagram of a three-level V-system. 
 The pump and probe fields have a frequency detunings
$\Delta_2$ and $\Delta_1$ respectively.  The $\gamma$'s denote the spontaneous
emission rates from the respective levels. (b) The arrangement of
field polarization required for single field driving one transition if
dipoles are non-orthogonal.}
\label{one}
\end{figure}
\noindent  The pump field 
($\vec{E}_2 = \vec{\varepsilon}_2 e^{-i\omega_2 t} + {\rm c.c}$) 
 with a Rabi frequency $2G = 2\vec{d}_{23}\cdot\vec{\varepsilon}_2/\hbar$ drives 
$|2\rangle \leftrightarrow |3\rangle$ transition ($\vec{d}_{13} \cdot 
\vec{\varepsilon}_2 = 0$) and similarly probe field ($\vec{E}_1 = \vec{\varepsilon}_1  
e^{-i\omega_1 t} + {\rm c.c}$) 
with a Rabi frequency $2g = 2\vec{d}_{13}\cdot\vec{\varepsilon}_1/\hbar$ drives 
$|1\rangle \leftrightarrow |3\rangle$ transition ($\vec{d}_{23} \cdot 
\vec{\varepsilon}_1 = 0$).  We work under the condition such that VIC is important.
However, we would like to keep the situation rather parallel to the usual case.
Thus we assume a specific geometry of the Fig. \ref{one}(b), so that the 
pump (probe) does not couple to $|1\rangle \leftrightarrow |3\rangle$ 
($|2\rangle \leftrightarrow |3\rangle$) transition.
The Hamiltonian for this system will be
\begin{eqnarray}
H &=& \hbar W_{13}|1\rangle \langle 1| + \hbar W_{23}|2\rangle \langle 2| - 
\hbar(G|2\rangle \langle 3| e^{-i\omega_2 t} \nonumber \\
& &+ g|1\rangle \langle 3| 
e^{-i\omega_1 t} + {\rm H.c}),
\label{ham1}
\end{eqnarray}
where $\hbar W_{i3}$ ($i = 1,2$) is the energy of the state $|i\rangle$ when measured with 
respect to state $|3\rangle$.  In the rotating wave approximation the 
density matrix equations with the inclusion
of all the decay terms will be
\begin{eqnarray}
\dot{\rho}_{11} &=& -2\gamma_1 \rho_{11} -\eta (\rho_{12}+\rho_{21}) +
ige^{-i\delta t}\rho_{31} -ig e^{i\delta t} \rho_{13},\nonumber\\
\dot{\rho}_{22} &=& -2\gamma_2\rho_{22} -\eta(\rho_{12}+\rho_{21}) +
iG\rho_{32} \nonumber\\
& & -iG\rho_{23},\nonumber\\
\dot{\rho}_{12} &=& -(\gamma_1 +\gamma_2 + iW_{12})\rho_{12} -\eta(\rho_{11} +  
\rho_{22}) \nonumber\\
& &+ i g e^{-i\delta t}\rho_{32} - i G \rho_{13},\nonumber\\
\dot{\rho}_{13} &=& -(\gamma_1 + i(\Delta_2 + W_{12}))\rho_{13} -\eta \rho_{23}
-iG\rho_{12} \nonumber\\
& &+ ige^{-i\delta t}(1 - 2\rho_{11} - \rho_{22}),\nonumber\\
\dot{\rho}_{23} &=& -(\gamma_2 + i\Delta_2)\rho_{23} -\eta\rho_{13} -
ige^{-i\delta t}\rho_{21} \nonumber\\
& &+ iG(1 - \rho_{11} - 2\rho_{22}),
\label{density1}
\end{eqnarray}
where $\delta = \omega_1 - \omega_2$ is the probe-pump detuning.
 The probe
detuning $\Delta_1 = W_{13} - \omega_1$ and the pump detuning $\Delta_2 = 
W_{23} - \omega_2$ are related by $\Delta_1 - \Delta_2 = W_{12} - \delta$.
In deriving (\ref{density1}) we have made the canonical transformations 
so that $\rho_{13}$ and 
$\rho_{23}$ are obtained by multiplying the solution of (\ref{density1}) 
by $e^{-i\omega_2 t}$.  We also use
the trace condition $\rho_{11} + \rho_{22} + \rho_{33} = 1$.
Here $\eta = \eta_0 \sqrt{\gamma_1 \gamma_2} \cos\theta$ is the VIC parameter, which 
is nonzero when  $\theta \neq 90^o$.  Note that for the 
 geometry shown in Fig. \ref{one}(b), $\theta$ is always nonzero, though it 
could be small. 
The parameter $\eta_0$ enable us to study the limiting case when the
effects of VIC are ignored ($\eta_0 = 0$), otherwise we will set $\eta_0 = 1$.
In the absence of external fields as seen from equations (\ref{density0}),
the VIC effect is important when the separation between the
two excited levels is of the order of natural line width.   However, this
condition may be relaxed when the system is being driven by external
fields as we will see later.

Let us first consider the case when $\eta_0 = 0$.  Making a further canonical 
transformation on $\rho_{13}$ and $\rho_{12}$ we can get
 rid of the explicit time dependence.  
The imaginary part of $\rho_{13}$ yields the probe absorption. In the limit
of a weak probe field ($g \ll \gamma_1, \gamma_2$), we obtain 
\end{multicols}
\hrule
\widetext
\begin{equation}
\rho_{13} = \frac{g\{(\gamma_2^2+\Delta_2^2+G^2)(\Delta_2 - \Delta_1 + 
i(\gamma_1+\gamma_2)) + G^2(\Delta_2 - i\gamma_2)\}}{(\gamma_2^2+\Delta_2^2+2G^2)
[G^2 + (\Delta_1 - i\gamma_1)(\Delta_2 - \Delta_1 +i(\gamma_1+\gamma_2))]}.
\end{equation}
\begin{multicols}{2}
\noindent
In the limit of vanishing $\gamma$'s and large $G$, the 
above expression shows that two complex poles exists at
$\Delta_1 = (\Delta_2+i\gamma_2 + 2i\gamma_1 \pm \sqrt{(\Delta_2 + 
i\gamma_2)^2 + 4G^2}~)/2$. The probe absorption
as a function of $\Delta_1$, i.e. as a function of probe frequency
 will have two resonances at $\Delta_1 = (\Delta_2 
\pm \sqrt{\Delta_2^2 + 4G^2})/2$).  These are  the two 
Autler-Townes components in the absorption spectrum.  It can be further shown
that ${\rm Im}(\rho_{13}) > 0$. 

We now consider the effects of VIC ($\eta_0 = 1$).  The system of Eqs. 
(\ref{density1}) have been studied under a very wide range of conditions.
We would now recall what has been done and in what ways our current work 
differs from the existing works.  (a) We could first consider the case when
pump is also replaced by the probe ($\vec{\varepsilon}_1 \equiv 
\vec{\varepsilon}_2$, $\omega_1 = \omega_2$). 
 Here the effects of VIC
manifest both in emission \cite{{plenio92},{swain1}} and absorption spectrum
\cite{{stroud82},{swain2},{paspalakis2}}.  Zhou and Swain 
demonstrated the existence of ultra-narrow spectral lines in emission 
\cite{swain1}.  Cardimona {\em et al.} showed vanishing of absorption under
certain conditions \cite{stroud82} whereas Zhou and Swain demonstrated the 
possibility of gain with no pump field present \cite{swain2}.   
(b)  Another case which is extensively studied by Knight and coworkers 
corresponds to degenerate pump and probe fields, i.e. $\vec{\varepsilon}_1 \neq
\vec{\varepsilon}_2$, but $\omega_1 = \omega_2$ ($\delta = 0$).  Here the pump
can have arbitrary strength  while the probe is kept relatively weak.  
Paspalakis {\em et al.} showed how VIC can lead to gain without inversion
\cite{paspalakis2}.  (c)  In the present work we study the important case of
non-degenerate pump and probe fields, $\vec{\varepsilon}_1 \neq \vec{\varepsilon}_2$, $\omega_1 \neq \omega_2$.  We show how the VIC can invert the traditional
Autler-Townes splittings in the absorption spectrum and produce gain features.

While we work with a three-level system, it should be noted that effects of VIC
in the context of four and five-level schemes have been very extensively investigated
\cite{{harris89},{zhu-scully96},{xia96},{agarwal97},{knight98-99},{zhu3},{piperno},{berman98},{anil98-99}}.
In particular, Zhu and coworkers discovered quenching
of spontaneous emission \cite{{zhu-scully96},{xia96}}.  An intuitive
picture for spontaneous emission suppression and enhancement was provided
by Agarwal \cite{agarwal97}.

 The non-degenerate case, that we treat, has a major complication due to 
explicit  
time dependence in the equations of
motion (\ref{density1}).  
Since the time dependence in (\ref{density1})
is periodic, we can solve  
these equations by Floquet analysis.  The solution can be
written as 
\begin{equation}
\rho_{ij} = \sum_m \rho^{(m)}_{ij}e^{-im\delta t}.
\label{sol1}
\end{equation}
Thus the absorption and emission spectra gets modulated at various
harmonics of $\delta$.  The dc component in probe absorption spectrum is
related to $\rho^{(+1)}_{13}$.  The absorption coefficient $\alpha$ per
unit length can be shown to be 
\begin{equation}
\alpha = \frac{\alpha_0 \gamma_1}{g}{\rm Im}(\rho^{(+1)}_{13}),
\label{sol2}
\end{equation}
where $\alpha_0 = 4\pi {\cal N}|d_{13}|^2\omega_1 /\hbar \gamma_1 c$ and ${\cal N}$ denote
the atomic density.  Note that in 
(\ref{sol2}) only one term from the entire series (\ref{sol1}) contributes.
For the case of degenerate pump-probe ($\delta = 0$), all the terms in the 
series (\ref{sol1}) are important.

\section{Numerical Results}
\begin{figure}[h]
\hspace*{-0.1 cm}
\epsfxsize 3 in
\epsfysize 4.4 in
\epsfbox{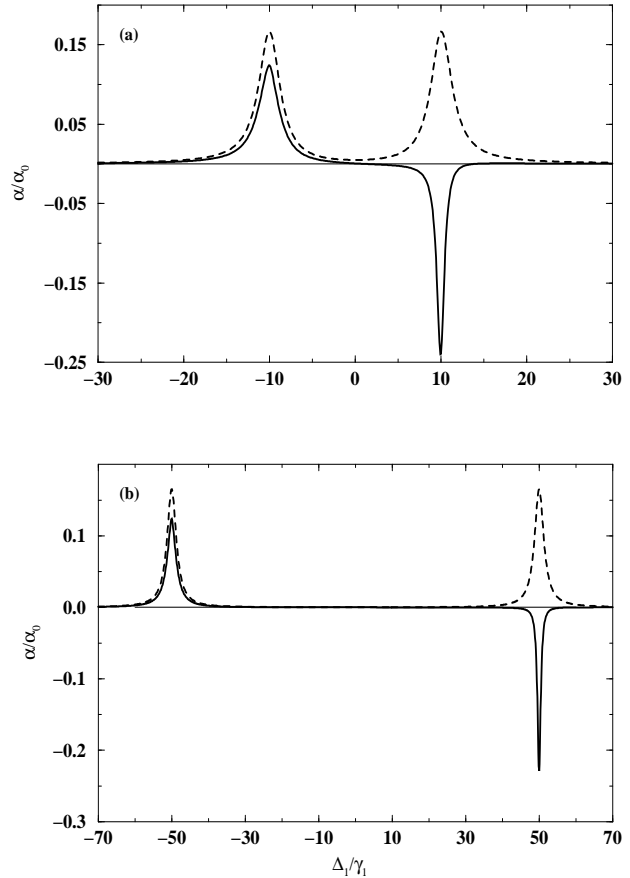}
\narrowtext
\caption{Effect of interference between decay channels on probe absorption.
For both the frames the dashed curves show the
usual Autler-Townes components in the absence of VIC ($\eta_0 = 0$) and the
solid curves is for $\eta_0 = 1$.  The common parameters are
$g = 0.01 \gamma_1$, $\gamma_2 = \gamma_1$, $\theta = 15^o$,
and $\Delta_2 = 0$.  Note that $\alpha$ will
depend on $W_{12}$ only when VIC is present and we take $W_{12} = -G$
when $\eta_0 = 1$. In frame
(a) we have kept $G = 10 \gamma_1$ and
in frame (b) we take $G = 50 \gamma_1$.
The solid curve in frame (b) shows that the effect of VIC is
retained even for large $W_{12}$.}
\label{two}
\end{figure}

\begin{figure}[h]
\hspace*{0.5 cm}
\epsfxsize 2.3 in
\epsfysize 2.1 in
\epsfbox[127 288 478 602]{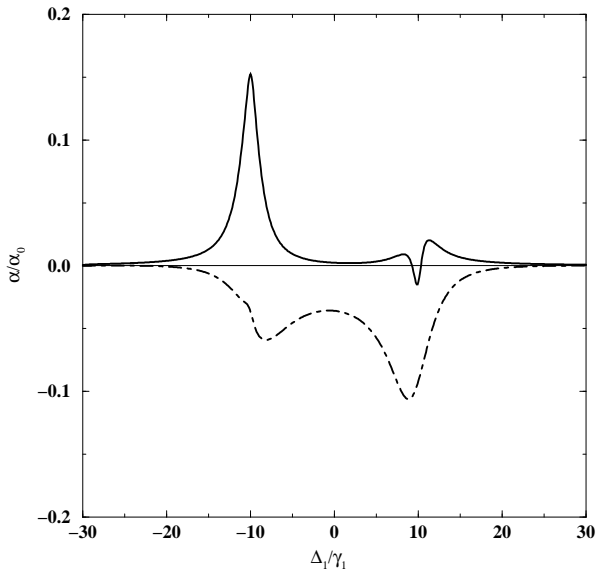}
\vspace*{2.31 cm}
\narrowtext
\caption{Plots show the important role $\theta$ and unequal $\gamma$'s play
in the presence of VIC ($\eta_0 = 1$).  The common parameters are $G = 10\gamma_1$, $g = 
0.01\gamma_1$, $\Delta_2 = 0$, and $W_{12} = -G$.  The solid curve presents the
case when $\gamma_2 = \gamma_1$ and $\theta = 35^o$ and the dot-dashed
curve arises when $\gamma_2 = 6\gamma_1$ and $\theta = 15^o$.}
\label{three}
\end{figure}

In order to obtain the probe absorption spectra we solve (\ref{density1}) 
numerically using the series solution (\ref{sol1}) and the steady state
condition $\dot{\rho}^{(m)}_{ij} = 0$.  The situation is
much simpler for a weak probe when $\rho^{(+1)}_{ij}$ can be computed to
first order in $g$, otherwise we use Floquet method.  
In Fig. \ref{two} we plot the probe absorption as a function of probe detuning.
The dashed curves in Fig. \ref{two}(a,b) are the usual Autler-Townes components 
in the absence of VIC effects.  The solid curves show the absorption spectra
when VIC is included.  We observe that {\em one of the Autler-Townes component flips 
sign to give rise to significant gain}.  This type of behavior is seen for
any value of $W_{12}$ provided the pump field strength satisfies the 
condition $G = |W_{12}|$  
\vskip 0.8 cm
\noindent
for $\Delta_2 = 0$.
When $W_{12} = G$ and $\Delta_2 = 0$, the gain appears at $\Delta_1 = -G$.
The solid curve in Fig. \ref{two}(b) shows that the effect of VIC is observed
even for large $W_{12}$ compared to $\gamma_1,\gamma_2$.  This is in contrast 
to the situation that exist in the absence of external fields where
one finds that VIC effects are important
when the separation between vacuum coupled levels is of the order of
natural line-width.
As can be seen from the Fig. \ref{two}(b), for strong pump fields,
such a restriction can be relaxed.
Also note that one of the Autler-Townes component can be almost
suppressed for certain set of parameters (see for example 
the solid curve in
Fig. \ref{three}).  Thus the
parameter $\theta$ (angle between the two transition dipole matrix elements)
 controls the spectra in presence of VIC.  The
dot-dashed curve in Fig. \ref{three} also shows the effect of unequal decays.
For
$\gamma_2 > 2\gamma_1$ both the Autler-Townes components flip. 
We analyze the origin of gain in the following sections.  We note that the previous works [17a] on the {\em degenerate} pump and probe fields also reported
gain, provided the energy separation between the two excited states can be
scanned.  

Finally as mentioned in Sec. II the observation of VIC related effects requires the use
of transitions with non-orthogonal dipole matrix elements \cite{{gsabook},{agassi84},{russian99},{plenio92},{swain1},{stroud82},{swain2},{paspalakis2},{zhu-scully96},{xia96},{agarwal97},{knight98-99},{zhu3},{piperno},{berman98},{anil98-99}}.  The question of production of transition with 
non-orthogonal dipole matrix elements has been extensively discussed in the literature.
This can be achieved by mixing the states using either internal fields 
\cite{xia96} or external fields \cite{{berman98},{anil98-99},{hakuta91},{russianbook}}.  We may further note that the relaxation need not occur by spontaneous
emission.  For example in problems involving intersubband transitions in 
semiconductors the relaxation can occur by emission of LO phonon \cite{imamoglu96}.  In such cases the non-orthogonality of dipole matrix elements is not required.

\section{Quasi-trapped-states from Interference of decay channels}
For a very weak probe field ($g \ll \gamma_1, \gamma_2$) we can
solve equations (\ref{density1}) perturbatively with respect to the
strength of the probe field. To the lowest order in $g$, the solution may 
be written as
\begin{equation}
\rho_{ij} = \sigma_{ij}^0 + g \sigma_{ij}^+ e^{-i\delta t} + g^*\sigma_{ij}^- 
e^{i\delta t}.
\label{sol3}
\end{equation}
We first examine the behavior of the system in the presence of pump 
field alone ($g = 0$).
Note that in the absence of VIC effects the system reduces to the well known
case of coherently driven two-level atom.  But the
behavior is quite different in the presence of VIC effects as we show in the
following.

\begin{figure}[h]
\hspace*{0.5 cm}
\epsfxsize 2.6 in
\epsfysize 4 in
\epsfbox[80 80 560 800]{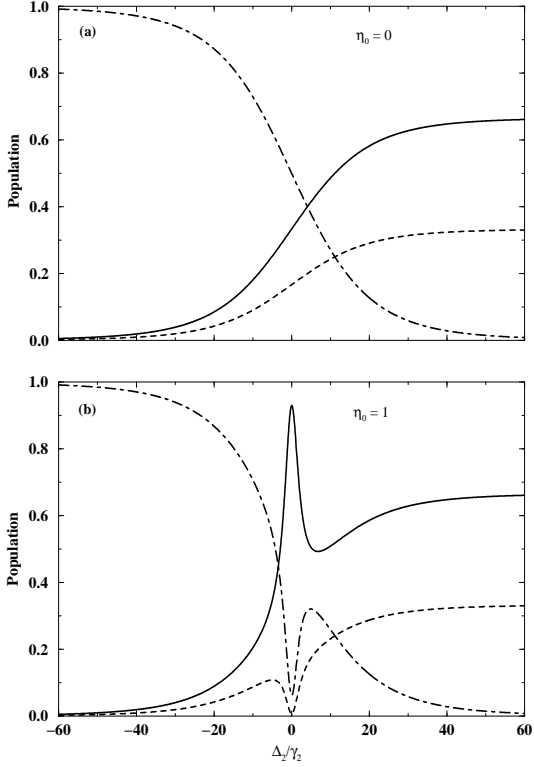}
\vspace*{0.5 cm}
\narrowtext
\caption{The atomic population in the basis (\ref{basis}) as a function of pump
detuning $\Delta_2/\gamma_2$ in presence (frame b) and in absence (frame a)
of VIC.  The parameters
are $G = 20 \gamma_2$, $W_{12} = -G$, $\gamma_1 = \gamma_2$ and $\theta = 15^o$.
  The solid
 curves denote $\sigma^0_{ucuc}$, the dashed curves are for $\sigma^0_{cc}$ and
the dot-dashed
curves denote $\sigma^0_{++}$.}
\label{four}
\end{figure}

It is clear that field $G$
creates a coherent mixing of states $|2\rangle$ and $|3\rangle$.  
The new eigenvalues will be
\begin{equation}
\lambda_{\pm} = \frac{\Delta_2 \pm \sqrt{\Delta_2^2 + 4G^2}}{2},
\end{equation}
 and the corresponding dressed energy states can be written as
\begin{eqnarray}
|+\rangle &=& \cos\psi |2\rangle + \sin\psi |3\rangle,\nonumber\\
|-\rangle &=& -\sin\psi|2\rangle + \cos\psi|3\rangle,
\label{dressb}
\end{eqnarray}
where $\tan\psi = -G/\lambda_+$.  The crucial point to note is that 
the level $|1\rangle$ is coupled with $|\pm\rangle$ because of the presence
of VIC.
Thus the population in $|\pm\rangle$ also depends on the VIC parameter $\eta$.  
An important case arises when $|1\rangle$ is degenerate with
either $|\pm\rangle$, i.e. when $W_{12} = \lambda_{\pm}$.  The degenerate
levels get strongly coupled via VIC, giving rise to trapping.  When $|1\rangle$
and $|-\rangle$ are degenerate, we show that
the dynamical behavior of the system can be best analyzed in the basis 
given below.
\begin{equation}
|+\rangle,~~~|c\rangle = \frac{\sqrt{2\gamma_1}|1\rangle + \sqrt{\gamma_2}|-\rangle}{\sqrt{\gamma_2+2\gamma_1}},
~~~|uc\rangle = \frac{\sqrt{\gamma_2}|1\rangle - \sqrt{2\gamma_1}|-\rangle}{\sqrt{\gamma_2+2\gamma_1}}.
\label{basis}
\end{equation}
 Using the transformations
(\ref{dressb}), (\ref{basis}) and Eqs. (\ref{density1}) with $g = 0$, we numerically
compute the steady state population in the states (\ref{basis}).  In Fig. 
\ref{four} we plot the population of these states as a function of pump 
detuning.
Note that in the presence of VIC, $\sigma^0_{ucuc}$ approaches
unity at $\Delta_2 = 0$ i.e. when the 
states $|1\rangle$ and $|-\rangle$ are degenerate because $W_{12} = -G$. 
A similar kind of trapping will occur when $|1\rangle$ is degenerate
with $|+\rangle$.  When
$|W_{12}| \neq G$ the trapping will occur for an off-resonant pump field.
Trapping also requires $\theta$ to be small.
We show later that $\sigma^0_{ucuc}$ cannot approach unity, and for this reason
we refer to it as `quasi-trapped-state' (QTS).
Figure \ref{four} also shows 
that for $\Delta_2 \ll -G$, all the population remains in
$|+\rangle$.  This is not an interference effect and happens irrespective
of whether VIC is present or absent.  For large negative pump detuning,
$\lambda_+ \rightarrow 0$ and thus
$\sin\psi \rightarrow 1$ ($\cos\psi \rightarrow 0$) in (\ref{dressb}).
Since the level $|3\rangle$ being the ground state, most
of the population remains here if the pump is highly off-resonant. 
The QTS $|uc\rangle$ is a result of
interference among decay channels of $|1\rangle$
and $|-\rangle$ levels.  As a consequence, even if $W_{12}$ is large in bare 
basis, strong
VIC effects can appear when dressed levels are degenerate with the bare excited
levels unconnected by the pump field.

\begin{figure}[h]
\hspace*{0.7 cm}
\epsfxsize 2.2 in
\epsfysize 2. in
\epsfbox[127 288 490 652]{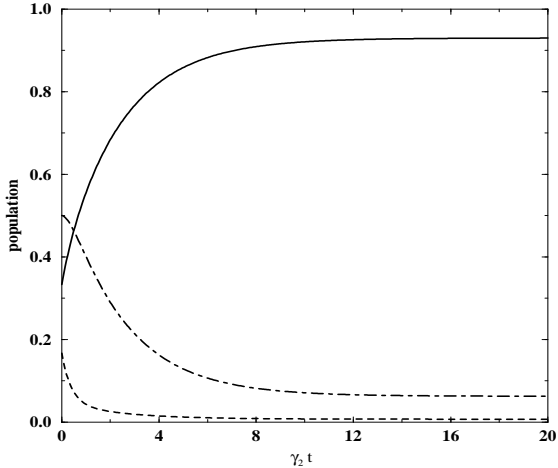}
\vspace*{0.7 cm}
\narrowtext
\caption{Evolution of atomic population in the states $|+\rangle$, $|c\rangle$,
and
$|uc\rangle$ when $\Delta_2 = 0$, $W_{12} = -G$ and the VIC parameter $\eta_0 =
1$.
The other parameters are
$G = 20\gamma_2$, $\gamma_1 = \gamma_2 = 1$, and $\theta = 15^o$. The solid
curve represents $\sigma^0_{ucuc}$, the dashed curve is for $\sigma^0_{cc}$ and
the dot-dashed curve shows evolution of $\sigma^0_{++}$.}
\label{five}
\end{figure}

We next examine how the quasi-trapped-state is formed.  For this purpose we
transform the
equations of motion (\ref{density1}) for the density matrix elements in 
basis (\ref{basis}).  For $\Delta_2 = 0$ and $W_{12} = -G$, a long calculation
leads to
\end{multicols}
\hrule
\widetext
\begin{mathletters}
\begin{eqnarray}
\dot{\sigma}^0_{ucuc} &=& -\frac{4\gamma_1\gamma_2(\gamma_1+\gamma_2)
(1-\cos\theta)}{(2\gamma_1+\gamma_2)^2}\sigma^0_{ucuc} + \frac{\gamma_1
(4\gamma_1^2 + 4\gamma_1\gamma_2\cos\theta + \gamma_2^2)}{(2\gamma_1
+\gamma_2)^2}\sigma^0_{cc} + \frac{\gamma_1\gamma_2}{2\gamma_1+\gamma_2}
\sigma^0_{++}\nonumber\\
& &-\frac{\gamma_2\sqrt{\gamma_1\gamma_2}(2\gamma_1 - \gamma_2)(1-\cos\theta)}
{\sqrt{2}(2\gamma_1+\gamma_2)^2}(\sigma^0_{ucc}+\sigma^0_{cuc}),
\label{11a}
\end{eqnarray}
\begin{eqnarray}
\dot{\sigma}^0_{cc} &=& -\frac{(4\gamma_1+\gamma_2)(4\gamma_1^2+4\gamma_1\gamma_2
\cos\theta +\gamma_2^2)}{2(2\gamma_1+\gamma_2)^2}\sigma^0_{cc} + \frac{2\gamma_1 
\gamma_2^2(1-\cos\theta)}{(2\gamma_1 +\gamma_2)^2}\sigma^0_{ucuc} + 
\frac{\gamma_2^2}{2(2\gamma_1+\gamma_2)}\sigma^0_{++}\nonumber\\
& &-\frac{\gamma_1(2\gamma_1-
\gamma_2)\sqrt{2\gamma_1\gamma_2}(1-\cos\theta)}{(2\gamma_1+\gamma_2)^2}
(\sigma^0_{ucc}+\sigma^0_{cuc}),
\end{eqnarray}
\begin{eqnarray}
\dot{\sigma}^0_{++} &=& -\frac{\gamma_2}{2}\sigma^0_{++} + \frac{(4\gamma_1^2+4\gamma_1
\gamma_2\cos\theta + \gamma_2^2)}{2(2\gamma_1+\gamma_2)}\sigma^0_{cc} + 
\frac{2\gamma_1\gamma_2(1-\cos\theta)}{(2\gamma_1+\gamma_2)}\sigma^0_{ucuc}\nonumber\\
& & +
 \frac{(2\gamma_1-\gamma_2)\sqrt{\gamma_1\gamma_2}(1-\cos\theta)}{\sqrt{2}
(2\gamma_1+\gamma_2)}(\sigma^0_{ucc} +\sigma^0_{cuc}),
\end{eqnarray}
\begin{eqnarray}
\dot{\sigma}^0_{ucc} &=& -\left[\frac{4\gamma_1^2\gamma_2(1-\cos\theta)}{(2\gamma_1+
\gamma_2)^2} + \frac{\gamma_1\gamma_2\cos\theta + \gamma_2^2}{(2\gamma_1+
\gamma_2)} + \gamma_1\right]\sigma^0_{ucc} - \frac{\gamma_1\gamma_2(1-\cos\theta)
(2\gamma_1-\gamma_2)}{(2\gamma_1+\gamma_2)^2}\sigma^0_{cuc}\nonumber\\
& & -\frac{\sqrt{\gamma_1
\gamma_2}(1-\cos\theta)}{\sqrt{2}}\sigma^0_{ucuc} - [8\gamma_1^2(1-\cos\theta) 
+ (2\gamma_1+\gamma_2)^2\cos\theta]\frac{\sqrt{\gamma_1\gamma_2}}{\sqrt{2}
(2\gamma_1 + \gamma_2)^2}\sigma^0_{cc} \nonumber\\
& &- \frac{\gamma_2\sqrt{\gamma_1\gamma_2}}
{\sqrt{2}(2\gamma_1 + \gamma_2)} \sigma^0_{++},
\end{eqnarray}
\begin{eqnarray}
\dot{\sigma}^0_{+uc} = -\frac{\gamma_2}{2(2\gamma_1+\gamma_2)}[\gamma_2 +
\gamma_1(1 + 6\sqrt{2} - \cos\theta)]\sigma^0_{+uc} - \frac{\sqrt{\gamma_1\gamma_2}}
{\sqrt{2}(2\gamma_1 + \gamma_2)}[2\gamma_1 + \gamma_2\cos\theta-2\gamma_2]
\sigma^0_{+c},
\end{eqnarray}
\begin{eqnarray}
\dot{\sigma}^0_{+c} = -\frac{\sqrt{2\gamma_1\gamma_2}}{(2\gamma_1+\gamma_2)}
[\gamma_1(1-\cos\theta) - \gamma_2]\sigma^0_{+uc} - [2\gamma_1\gamma_2(1+\cos\theta)
+ 4\gamma_1^2 + 3\gamma_2^2]\frac{\sigma^0_{+c}}{2(2\gamma_1 + \gamma_2)}.
\end{eqnarray}
\end{mathletters}
\begin{multicols}{2}
The above equations have been derived by neglecting terms rotating at
$e^{\pm 2iG t}$ (secular approximation).  Note that the above equations are
not the usual rate equations because the diagonal elements are coupled
with the off-diagonal elements as in (\ref{density0}).  It is this coupling
which leads to quasi-trapping even though 
$\sigma^0_{ucuc}$ decays at a rate 
\begin{equation}
\Gamma_{uc} = \frac{4\gamma_1\gamma_2(\gamma_1+\gamma_2)(1-\cos\theta)}
{(2\gamma_1 + \gamma_2)^2}.
\end{equation} 
Also note that for small non-zero $\theta$ the decay from state $|uc\rangle$
is very small, which makes it a highly `stable' state.  We solve Eqs. (11)
numerically with the initial condition $\sigma^0_{33}(0) = 1$.  In Fig. \ref{five}
we plot the time evolution of the population in states $|+\rangle$, 
$|c\rangle$, and $|uc\rangle$.  Note that both $|+\rangle$ and $|c\rangle$
decay very rapidly, while population gets accumulated in $|uc\rangle$. Here  
complete trapping will occur
($\sigma^0_{ucuc}(\infty) = 1$) when $\Gamma_{uc} = 0$.  This is not possible
for the geometry shown in Fig. \ref{one}(b).  However, we have an  
quasi-trapped-state for small $\theta$. 

It should be noted that trapped states were shown to occur in 
presence of VIC under several conditions.
For example trapped state arises
at certain parameter regime when $|1\rangle$ and $|2\rangle$ are degenerate
and when no external fields are applied, however 
the system is prepared in one of the
excited states \cite{gsabook}.  Trapping is also known to 
occur in the degenerate case ($\delta = 0$) and when pump and probe have
identical strengths ($\vec{\varepsilon}_2 \equiv \vec{\varepsilon}_1$)
\cite{swain1}.  Recently, new trapping states in the presence 
of VIC have been found for four-level systems \cite{{zhu3},{piperno}}.
However, note that the QTS discussed above is due to 
a pump field $G$ coupling $|2\rangle \leftrightarrow |3\rangle$
transition and thus is {\em different} from all the previous works.  

\vskip 0.5 cm

\section{Effects of Quasi-trapped-state on the pump field line-profiles}
\begin{figure}[h]
\hspace*{0.7 cm}
\epsfxsize 2.3 in
\epsfysize 3.6 in
\epsfbox[85 120 520 800]{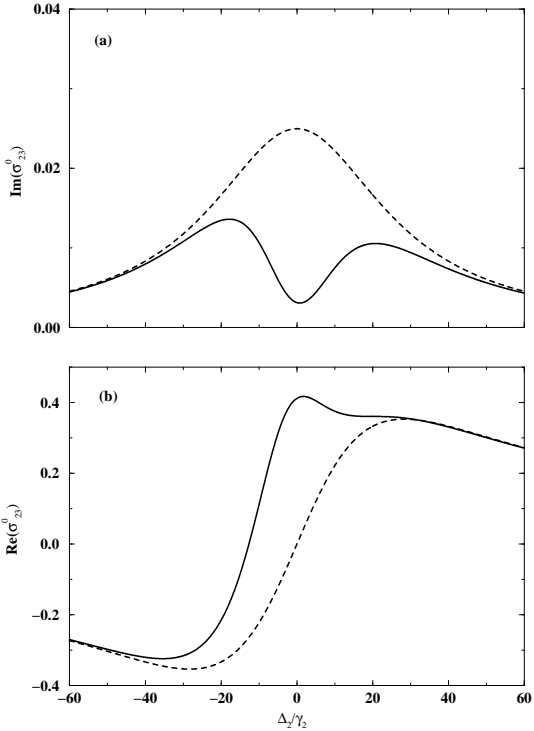}
\narrowtext
\vspace*{1.5 cm}
\caption{Plots show the absorption and dispersion curves for the pump field in
dimensionless units as a function of pump detuning $\Delta_2/\gamma_2$.
The solid curves show the effect of 
VIC and the dashed curves are for $\eta_0 = 0$.  The parameters 
are $G = 20\gamma_2$, $\gamma_1 = 10\gamma_2$, and $\theta = 15^o$.  
For the solid curve
$W_{12} = -G$.}
\label{six}
\end{figure}

We now show the effects of the above trapping on the absorption and
dispersion profiles of the pump field.  The trapping leads to a very steep 
increase in the refractive index and reduces absorption drastically for the
pump field.  Various models in the past have demonstrated and discussed
the importance of such a medium \cite{{scully2},{olga1},{zibrov96}}.
However, in the present case we show how VIC can be used to control the 
refractive index of a medium.  It is
known that a large population difference between the dressed states can 
result in large dispersion with vanishing absorption \cite{olga1}.
For our system the population of dressed states depends on $\eta$ and 
 hence in principle we can get a situation where large population difference
between dressed states can exist.  
Consider the case when 
$\Delta_2 = 0$
and $W_{12} = -G$.  The coherence $\sigma^0_{23}$ can be evaluated using Eqs.
(\ref{density1}) and (\ref{sol3}).  The optical coherence to all orders in the
pump field is found to be 
\begin{eqnarray}
\sigma^0_{23} &=& [G^2 \eta^2\{G^2(2\gamma_1+\gamma_2)\gamma_1 + (\eta^2 -\gamma_1\gamma_2)(\gamma_1+\gamma_2)^2\}\nonumber\\
& &+ iG(\gamma_1\gamma_2-\eta^2)\{A\gamma_2 - \eta^2\gamma_1(\gamma_1+\gamma_2)^2\}]/B,
\label{rho231}
\end{eqnarray}
where 
\begin{eqnarray}
A &=& G^2\gamma_2^2 + 4G^2\gamma_1\gamma_2 + \gamma_1^2\gamma_2^2 + 4G^2\gamma_1^2 + 2\gamma_1^3\gamma_2 + \gamma_1^4,\nonumber\\
B &=& (\gamma_1\gamma_2-\eta^2)\{A(\gamma_2^2+2G^2) + \eta^2 G^2 \gamma_1(\gamma_2+2\gamma_1)\} \nonumber\\
& &+ \eta^2G^4 (\gamma_2+2\gamma_1)^2 
+ \eta^2
(\gamma_1+\gamma_2)^2(3\gamma_1\gamma_2\eta^2 \nonumber\\
& &- 2\gamma_1^2\gamma_2^2 - \eta^4).
\end{eqnarray}
It is known that the ${\rm Re}(\sigma^0_{23})$ corresponds to the dispersion
and ${\rm Im}(\sigma^0_{23})$ corresponds to absorption. 
When the alignment parameter $\theta$ is small, 
we have $\eta^2 \approx \gamma_1\gamma_2$ (for example when $\theta = 15^o$,
$\eta^2 = 0.93\gamma_1\gamma_2$), then we can approximate (\ref{rho231}) by 

\begin{equation}
\sigma^0_{23} \approx \frac{\gamma_1}{\gamma_2+2\gamma_1} +i\frac{(\gamma_1
\gamma_2-\eta^2)\{A - \gamma_1^2(\gamma_1+\gamma_2)^2\}}
{G^3\gamma_1(\gamma_2+2\gamma_1)^2},
\label{rho232}
\end{equation}
with the constraint that $G \neq 0$.
Thus for $\gamma_1 > \gamma_2$ one can have ${\rm Re}(\sigma^0_{23})$ as high as
0.5 while the absorption remains low.  It should be borne in mind that the absorption
and dispersion have been computed to all orders in the pump field strength.
In Fig. \ref{six} we plot the absorption and dispersion
parts of $\sigma^0_{23}$ as a function of detuning $\Delta_2/\gamma_2$. 
For comparison the dashed curves show the result in the absence of VIC.
These curves are obtained from the steady state numerical solutions of (\ref{density1})
with $g = 0$. 
Note that in the presence of VIC there is a dip in absorption
and a peak in dispersion curve.  
Due to trapping most of the population tend to remain in states $|1\rangle$ and
$|-\rangle$.   
For $W_{12} = -G$ and $\Delta_2 = 0$ the population in the three
dressed states and the coherence $\sigma^0_{+-}$ was evaluated to be
\end{multicols}
\widetext
\begin{equation}
\sigma^0_{11} = \frac{G^2\eta^2(\gamma_1+\gamma_2)^2(\eta^2-\gamma_1\gamma_2) - G^4 \eta^2\gamma_2 (\gamma_2+2\gamma_1)}{B},
\label{rho11}
\end{equation}

\begin{eqnarray}
\sigma^0_{--} &=& [(\gamma_2^2+2G^2)A(\gamma_1\gamma_2-\eta^2)-\eta^2G^2
(\gamma_1\gamma_2-\eta^2)(3\gamma_2^2+5\gamma_1\gamma_2+\gamma_1^2)
+ 4G^4\eta^2\gamma_1\gamma_2 \nonumber\\
& &+ \eta^2(\gamma_2+\gamma_1)^2(3\eta^2\gamma_1\gamma_2 -
2\gamma_1^2\gamma_2^2 - \eta^4)]/2B,
\label{rho12}
\end{eqnarray}
\begin{equation}
\sigma^0_{++} = \frac{(\gamma_1\gamma_2-\eta^2)\{A(\gamma_2^2+2G^2)+
G^2\eta^2\gamma_1(\gamma_2+2\gamma_1)+\eta^2(\gamma_1+\gamma_2)^2(G^2 - 2\gamma_
1\gamma_2 + \eta^2)\}}{2B}.
\label{rho13}
\end{equation}
\begin{eqnarray}
\sigma^0_{+-} &=& [(\eta^2-\gamma_1\gamma_2)(A\gamma_2^2 - 
G^2(\gamma_2^2+\gamma_1\gamma_2 -\gamma_1^2)\eta^2) - \eta^2(\gamma_1+\gamma_2)^2
(3\gamma_1\gamma_2\eta^2-2\gamma_1^2\gamma_2^2 -\eta^4) \nonumber\\
& & - 2iG(\eta^2 - 
\gamma_1\gamma_2)(A\gamma_2 - \eta^2\gamma_1(\gamma_1+\gamma_2)^2)]/2B,
\label{rho14}
\end{eqnarray}
which under the condition $\eta^2 \approx \gamma_1\gamma_2$ reduce to
\begin{equation}
\sigma^0_{11} \approx \frac{\gamma_2}{\gamma_2+2\gamma_1}, ~~~~ \sigma^0_{--} \approx 
\frac{2\gamma_1}{\gamma_2+2\gamma_1},
\label{dress4}
\end{equation}
\begin{equation}
\sigma^0_{++} \approx \frac{(\gamma_1\gamma_2 -\eta^2)\{A(\gamma_2^2+2G^2)+
G^2\gamma_1^2\gamma_2(\gamma_2+2\gamma_1)+\gamma_1\gamma_2(\gamma_1
+\gamma_2)^2(G^2-\gamma_1\gamma_2)\}}{2G^4\gamma_1\gamma_2(\gamma_2+2\gamma_1)^2},
\label{dress41}
\end{equation}
\begin{equation}
{\rm Im}(\sigma^0_{+-}) \approx \frac{(\gamma_1\gamma_2-\eta^2)\{A-\gamma_1^2(\gamma_1
+\gamma_2)^2\}}{G^3\gamma_1(\gamma_2+2\gamma_1)^2}.
\label{dress5}
\end{equation}
\hrule
\begin{multicols}{2} 
\noindent
Note that $\sigma^0_{++}$ and ${\rm Im}(\sigma^0_{+-})$ are very small compared
to $\sigma^0_{11}$ and $\sigma^0_{--}$.
One can
equally write $\sigma^0_{23}$ as
\begin{equation}
\sigma^0_{23} = (\sigma^0_{--} - \sigma^0_{++})/2 +  i {\rm Im}(\sigma^0_{+-}),~~~~~~{\rm at \Delta_2 = 0.}
\label{rho233}
\end{equation}
Thus the {\em large difference in population between states
$|-\rangle$ and $|+\rangle$ gives rise to the large dispersion}.
  Also note that state
$|-\rangle$ lies below the ground state $|3\rangle$ which can also 
cause
large index of refraction with small absorption \cite{olga1}.
When $\eta_0 = 0$, we see from (\ref{rho12}), (\ref{rho13}) and (\ref{rho14})
that, 
$\sigma^0_{--} = \sigma^0_{++}$ and ${\rm Im}(\sigma^0_{+-}) \neq 0$, 
and hence dispersion at $\Delta_2 = 0$ is zero with substantial absorption
which is consistent with the   
well known power broadened absorption and dispersion profiles for a two-level
atom.

\section{Origin of Gain through Quasi-trapped-states}
The origin of the Autler-Townes doublet in the absorption spectrum is well 
understood.  The pump dresses the states $|2\rangle$ and $|3\rangle$.  The
population in the dressed states $|\pm\rangle$ absorbs a photon from the probe
field leading to the Autler-Townes doublet.  The situation changes drastically
in presence of VIC which as shown in Sec. V can, for a suitable choice of 
parameters, lead to a quasi-trapped-state $|uc\rangle$.
For $\Delta_2 = 0$, $W_{12} = -G$, $\gamma_1 = \gamma_2$ and small values of 
$\theta$ the dressed state $|+\rangle$ is almost empty where as $\sigma^0_{--}
> \sigma^0_{11}$ (Eq. (\ref{dress4})).  Thus the probe can be absorbed in the 
transition $|-\rangle \rightarrow |1\rangle$ whereas the probe will 
experience gain in the transition $|1\rangle \rightarrow |+\rangle$.  We also
note that in principle the coherence between two dressed states $|\pm\rangle$
can also contribute to the gain \cite{gsa91_Rapid}.  As discussed in the 
Sec V, the population in the states $|\pm\rangle$ and $|1\rangle$ depends on 
the angle $\theta$ between the two dipole matrix elements. 
For intermediate values of $\theta$'s the population in $|1\rangle$ and $|+\rangle$ can be
almost same.  This can suppress one of the Autler-Townes component as shown by the
solid curve in Fig. \ref{three}.  When $\gamma_2 > 2\gamma_1$, 
it is 
 possible to have $\sigma^0_{11} > \sigma^0_{--}$ for small $\theta$ (see result
(\ref{dress4})).  Thus both the Autler-Townes components will
show gain.  This behavior is shown by the dot-dashed curve in 
Fig. \ref{three}.   

\section{conclusions}
 
In summary, we have studied the {\bf non-degenerate} pump-probe spectroscopy of 
V-systems when the presence of interference in decay channels is significant.
We have shown the possibility of gain components in Autler-Townes doublet.
We present physical interpretation of this gain.  We have also shown the 
possibility of a new trapped states due to VIC which we further show,  
results in very high refractive index with very low absorption.  

One of us (GSA) thanks Olga Kocharovskaya and Marlan O. Scully for a 
number of discussions on the subject of vacuum induced coherences.

\end{multicols}
\end{document}